\documentclass[prl,twocolumn,amsmath,amssymb,superscriptaddress]{revtex4-2}

\usepackage{bm}                     
\usepackage{appendix}
\usepackage[latin1]{inputenc}
\usepackage{dcolumn}      
\usepackage{bbm}
\usepackage{color}
\usepackage{xcolor}

\usepackage[mathscr]{eucal}
\usepackage{latexsym}
\usepackage{amsbsy}
\usepackage{float}
\usepackage{graphicx}
\usepackage{epsfig}
\usepackage{subfigure}
\usepackage{wrapfig}
\usepackage{epsf}
\usepackage{float}
\usepackage[normalem]{ulem}
\usepackage{qcircuit}

\definecolor{darkblue}{HTML}{004D6B}
\definecolor{darkred}{HTML}{8c1515}
\definecolor{darkgreen}{HTML}{006400}
\usepackage{hyperref}
\hypersetup{ colorlinks=true, urlcolor=darkblue, citecolor=darkred,
    linkcolor=darkblue, breaklinks }

\newcommand{\ba}{\begin{array}}
\newcommand{\ea}{\end{array}}
\newcommand{\be}{\begin{equation}}
\newcommand{\ee}{\end{equation}}
\newcommand{\bea}{\begin{eqnarray}}
\newcommand{\eea}{\end{eqnarray}}

\newcommand{\new}[1]{\textcolor{black}{#1}}
\begin{document}

\title{Approaching ideal rectification in superconducting diodes through multiple Andreev reflections}

\author{A. Zazunov}
\affiliation{Institut f\"ur Theoretische Physik, Heinrich-Heine-Universit\"at, D-40225  D\"usseldorf, Germany}
\author{J. Rech}
\affiliation{Aix Marseille Univ., Universit\'e de Toulon, CNRS, CPT, Marseille, France}
\author{T. Jonckheere}
\affiliation{Aix Marseille Univ., Universit\'e de Toulon, CNRS, CPT, Marseille, France}
\author{B. Gr{\'e}maud}
\affiliation{Aix Marseille Univ., Universit\'e de Toulon, CNRS, CPT, Marseille, France}
\author{T. Martin}
\affiliation{Aix Marseille Univ., Universit\'e de Toulon, CNRS, CPT, Marseille, France}
\author{R. Egger}
\affiliation{Institut f\"ur Theoretische Physik, Heinrich-Heine-Universit\"at, D-40225  D\"usseldorf, Germany}

\date{\today}

\begin{abstract}
We analyze the rectification properties of voltage-biased Josephson junctions exhibiting the superconducting diode effect. Taking into account multiple Andreev reflection (MAR) processes in our scattering theory, we consider a
short weak link of arbitrary transparency between \new{two 
superconductors} with finite
Cooper pair momentum $2q$.  In equilibrium, the diode efficiency is bounded from above in this 
model, with maximal efficiency $\eta_0\approx 0.4$. Out of equilibrium, we find a rich subharmonic structure in the current-voltage curve.  For high transparency and 
low bias voltage $V$, the rectification efficiency $\eta(V)$ approaches the ideal value $\eta=1$ for
$q\xi\to 1$ (with coherence length $\xi$). 
\end{abstract}
\maketitle

\emph{Introduction.---}Starting with the work by Ando \emph{et al.}~in 2020 \cite{Ando2020}, 
a surge of experiments reported evidence for the superconducting diode effect (SDE)  \cite{Lyu2021,Bauriedl2022,Baumgartner2022,Pal2022,Lin2022,Wu2022a,Jeon2022,Turini2022,Sundaresh2023,Mazur2023,Anwar2023,Banerjee2023,Ghosh2023,Hou2023a,Costa2023}, 
see Ref.~\cite{Nadeem2023} for a review.  Even though microscopic mechanisms behind the SDE are not yet 
fully understood, many aspects have been clarified by theoretical works   \cite{Edelstein1996,Hu2007,Reynoso2008,Zazunov2009,Misaki2021,Ilic2022,He2022,Zhang2022,Davydova2022,Kokkeler2022,Daido2022a,Daido2022b,Tanaka2022,Zinkl2022,Halterman2022,Cheng2023,Ikeda2023,He2023a,Lu2023,Fu2023,Wang2023a,Yuan2023b,Legg2023,Nakamura2023,Picoli2023}, and the hope is that practically useful device applications will emerge soon. In essence, the SDE amounts to an asymmetry between the (absolute value of the) 
critical supercurrent flowing to the right ($I_{c+}>0$) and to the left ($I_{c-}<0$).
Assuming $|I_{c-}|<I_{c+}$, the
SDE efficiency is defined by $\eta_0 =(I_{c+}+I_{c-})/(I_{c+}-I_{c-})$,
where a dissipationless supercurrent $I$ \new{with $|I_{c-}|<|I|<I_{c+}$  can only flow to the right}. 
We consider the intrinsic SDE in a single Josephson junction (the SDE is also possible in junction-free bulk superconductors, see, e.g., Refs.~\cite{Ando2020,Bauriedl2022,Nadeem2023}, and
in more complex multi-junction devices \cite{Souto2022,Fominov2022,Paolucci2023,Gupta2023,Ciaccia2023,Zhang2023expA})
 \new{exhibiting the anomalous Josephson effect 
\cite{Buzdin2008,Reynoso2008,Zazunov2009,Reynoso2012,Brunetti2013,Dolcini2015,Campagnano2015,Szombati2016,Qin2017,Minutillo2018,Alidoust2021}, with 
broken time-reversal and inversion symmetries.  In addition, 
the current-phase relation (CPR) must contain higher harmonics} \cite{Reynoso2008,Zazunov2009,Bauriedl2022,Baumgartner2022}.

We here study the out-of-equilibrium behavior of
Josephson junctions exhibiting the SDE in equilibrium, 
in particular,  the DC current-voltage ($I$-$V$) curve of a voltage-biased
intrinsic Josephson diode.  At low temperatures,  
MAR processes \cite{Klapwijk1982,Bratus1995,Averin1995,Cuevas1996} 
can then provide the dominant transport mechanism, especially for subgap voltages $e|V|<2\Delta$
(with pairing gap $\Delta$).  We focus on junctions with a single (or a few uncoupled) 
channels, where the impedance is of order $h/e^2$ and thus much larger than the typical
impedance of the external circuit. We then do not have to account for
the self-consistent dynamics of the phase difference and voltage across the junction. 
Previous studies of nonequilibrium transport in Josephson diodes have considered weakly damped 
low-impedance junctions \cite{Misaki2021,Fominov2022,Trahms2023,Steiner2023} 
or externally driven junctions \cite{Paaske2023}, but MAR effects have not been addressed.
Our theory predicts a characteristic voltage-dependent 
rectification pattern, $I(-V)\ne -I(V)$, quantified by the efficiency parameter
\begin{equation}\label{efficiency}
    \eta(V) = \frac{I(V)+I(-V)}{I(V)-I(-V)},
\end{equation}
where $\eta(V)$ is especially large in the subgap regime. 
For the conventional case without SDE, MAR causes a subharmonic structure,
i.e., singular features in the nonlinear conductance for $eV=2\Delta/n$ with integer $n$ \cite{Klapwijk1982,Bratus1995,Averin1995}.   
For Josephson diodes, we predict an even richer subharmonic structure which determines
the rectification characteristics and might provide precious information about 
the microscopic mechanisms generating the SDE. 
Our central finding is that the efficiency $\eta(V)$ can approach the ideal limit of full rectification with $\eta=1$
at low voltages, even though $\eta_0\alt 0.4$ for the SDE efficiency in equilibrium for the model 
considered below. \new{The importance of MAR processes 
for rectification is related to the fact that higher harmonics of the CPR are 
needed for the SDE. Indeed, equilibrium Andreev states are the result of resonant MAR processes.  
A finite voltage breaks up the resonant MAR loop (see below) 
and effectively opens the way to high anharmonicity with many harmonics.  
In contrast to the current-biased case \cite{Davydova2022}, 
we find that the voltage-biased junction has a different optimal working point
and allows for ideal rectification.}
 
It is well known that the SDE can arise from magnetochiral effects \cite{Rikken2001,Tokura2018,Morimoto2018,Legg2022} 
in noncentrosymmetric superconductors \cite{Edelstein1996}. 
An alternative mechanism arises from the finite Cooper pair
 momentum $2q$ in a helical superconductor \cite{Davydova2022,Yuan2022,Pal2022,Lin2022,Yuan2022,Banerjee2023}.
We consider a weak link connecting two helical \new{superconductors, 
where the CPR computed in \cite{Davydova2022} implies the SDE. 
Scattering theory yields the exact $I$-$V$ curve for this model, where
known results \cite{Bratus1995,Averin1995,Zazunov2006} are recovered for $q=0$.    
A technical account is given in \cite{PRB}, where we also address the SDE efficiency $\eta_0$ in equilibrium.
 We here summarize the salient features of the theory and discuss the Josephson diode efficiency $\eta(V)$.}
 
\emph{Model.---}For \new{finite $q$},
the order parameter of an $s$-wave BCS superconductor oscillates in space, 
$\Delta(x)=\Delta e^{2iqx}$,
where $q\ne 0$ may originate from the interplay of the spin-orbit interaction with
a Zeeman field in superconducting films \cite{Daido2022a,He2022,Yuan2022,Levichev2023},
or from magnetic proximity and/or Meissner effects \cite{Davydova2022}. 
In either case, time-reversal and inversion symmetries are broken for $q\ne 0$.  
Following Ref.~\cite{Davydova2022}, we study a short single-channel weak link between two superconducting banks with the 
same pairing gap and Cooper pair momentum. The coherence length is $\xi=\hbar v_F/\Delta$ with Fermi velocity $v_F$.
For definiteness, we assume $0 \le q \xi < 1$ since the superconductor becomes 
gapless for $q\xi \ge 1$.  

Linearizing the band dispersion around the Fermi momentum points $\pm k_F$ with $k_F\xi\gg 1$, 
the Hamiltonian is expressed in terms of effectively one-dimensional quasiclassical Nambu spinor envelopes,
$\psi_{\pm}(x,t)=(\psi^{}_{\pm,\uparrow},\psi^\dagger_{\pm,\downarrow})^T$,
for right- and left-movers having momenta $\pm k_F+k$ with $|k|\ll k_F$,
resp., where $x<0$ ($x>0$) refers to the left (right) superconducting bank.
Gauging away the $e^{2iqx}$ factor from the order parameter, 
the Bogoliubov-de~Gennes (BdG) Hamiltonian for $x\ne 0$ follows as (we often put $\hbar= 1$) \cite{Davydova2022}
\begin{equation}\label{BdG}
H_{\rm BdG}  =  -iv_F \sigma_z\tau_z \partial_x + v_F q \sigma_z\tau_0 + \Delta\sigma_0\tau_x,
\end{equation} 
where we use Pauli matrices $\tau_{x,y,z}$ (and identity $\tau_0$) in Nambu space
and $\sigma_{x,y,z,0}$ in chiral (right-left mover) space.
Defining the bispinor $\Psi(x,t)=(\psi_+,\psi_-)^T$ in chiral space,
modeling the weak link as normal-conducting constriction with length much shorter than $\xi$ and 
transmission probability ${\cal T}$, and using the phase difference $\varphi(t)=2eVt$ 
across the junction, we arrive at a matching condition 
connecting the bispinors on the left ($x=0^-$) and right ($x=0^+$) side, see also \cite{Zazunov2005,Nazarov2009,Zazunov2014,Ackermann2023},
\begin{equation}\label{BC}
\Psi(0^-,t) = \frac{1}{\sqrt{\cal T}} (\sigma_0+r\sigma_x) \,  e^{i\tau_z eV t} \, \Psi(0^+,t),
\end{equation}
with the reflection amplitude $r=\sqrt{1-{\cal T}}$.  

\begin{figure}
\includegraphics[width=0.49\textwidth]{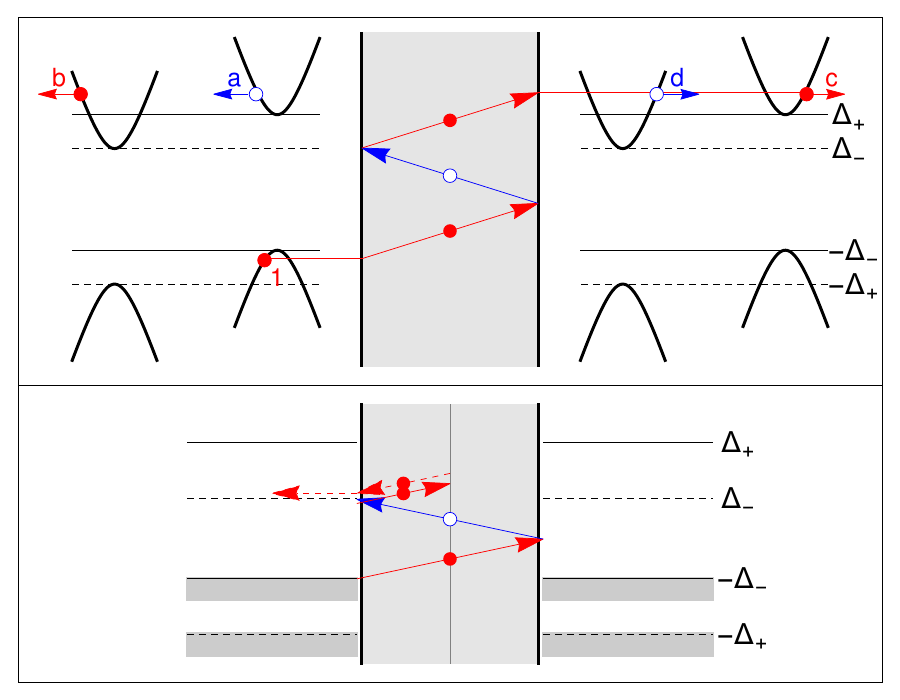} 
\caption{Sketch of the BdG dispersion and the MAR ladder picture.
Upper panel:  The dispersion \eqref{bulkenergy} is shown on the 
left and on the right side, where the shaded central region 
refers to the normal weak link. 
Electron $(e)$ and hole $(h)$ states are indicated by filled red (open blue) 
circles and their trajectories in energy space are shown as red (blue) lines.  
In this representation, the voltage drops across the normal region and 
the shown $e/h$ trajectories gain energy $eV$ by traversing it.
A typical MAR trajectory for ${\cal T}=1$ is shown, where an
$e$ state impinges from the left side (channel type $s=1$) with energy $E\alt -\Delta_-$. The scattering amplitudes $(a,b,c,d)$ in Eq.~\eqref{MARAnsatz} are also indicated. 
After Andreev reflection of the electron at the right (NS) interface, the 
reflected hole moves to the left side and gains energy $eV$ 
before being Andreev reflected to an electron at the SN interface.  
The final energy of the electron entering the right superconductor is $E_3=E+3eV$.  
For $E_3\approx \Delta_+$, the peak structure of the superconducting density of states implies a MAR feature, which here happens for $2\Delta/eV=3$.  Lower panel:
In the presence of normal reflection $r\ne 0$, i.e., for ${\cal T}<1$, new types of trajectories
are possible which generate additional MAR features for $2\Delta_\pm/eV=n$ (integer $n$).
In the shown example, the voltage is $eV=\Delta_-$. In the last step, the 
electron undergoes normal reflection and ends in the left side at 
energy $E\approx \Delta_-$.  This trajectory is of first order in $r$ and causes a MAR feature at $2\Delta_-/eV=2$.
}
\label{fig1}
\end{figure}

\emph{Spectral properties.---}Consider first a ballistic junction (${\cal T}=1$) at zero voltage, where Eq.~\eqref{BC} is automatically fulfilled by continuous wave functions
and one recovers the ``bulk'' case. BdG eigenstates have conserved energy $E$ and chirality
$\alpha=\sigma_z=\pm$, where Eq.~\eqref{BdG} gives a quasiparticle dispersion with four branches in total, see Fig.~\ref{fig1},
\begin{equation}\label{bulkenergy}
  E(k) = \pm \sqrt{(v_F k)^2+ \Delta^2}  + \alpha v_F q.
\end{equation} 
From Eq.~\eqref{bulkenergy}, there are \emph{two} positive threshold energies for the quasiparticle continuum, $\Delta_\pm=\Delta\pm v_Fq$, which are Doppler shifted away from $\Delta$ due to the finite Cooper pair momentum.  

For transparency ${\cal T}<1$, Andreev bound state solutions localized near the junction at $x=0$ can then only exist for energies inside both spectral gaps, $|E|<\Delta_-$.
On the other hand, for $|E|>\Delta_+$, propagating continuum states are possible along both directions, while for $\Delta_-<|E| <\Delta_+$, we have mixed-character states which can 
freely propagate along one direction but are evanescent along the other. 
As an important technical step forward, we specify BdG solutions that apply 
in a unified manner to all three energy regions.  We choose a formulation that can be 
leveraged to describe scattering states for finite voltage.
We first observe that for $x\ne 0$, using the Doppler-shifted energy $E_\alpha= E-\alpha v_F q$ for an $\alpha$-mover ($\alpha=\pm$) with energy $E$, 
Eq.~\eqref{BdG} implies that  electron ($e$) and hole ($h$) type states have the Nambu spinor structure 
\begin{equation}\label{tildepsi}
\tilde \psi_{\alpha,e}(E) = \frac{1}{\sqrt2}\left(\begin{array}{c} 1 \\ \rho(E_\alpha) \end{array}\right),
\quad \tilde \psi_{\alpha,h}(E) =\tau_x \tilde\psi_{\alpha,e}(E), 
\end{equation}
with the Andreev reflection amplitude
\begin{equation}\label{tildegamma} 
\rho(E)=  \left\{\begin{array}{cc}  
{\rm sgn}(E)\, \frac{|E|- \sqrt{E^2-\Delta^2}}{\Delta}, & |E|\ge \Delta,\\ & \\
\frac{E-i \sqrt{\Delta^2-E^2}}{\Delta}, & |E| <\Delta.
\end{array} \right. 
\end{equation} 
The states \eqref{tildepsi} describe arbitrary energies and are very useful for the description of outgoing (scattered) states, even 
though they satisfy unconventional normalization conditions.  On the other hand,
incident states, $\psi_{\alpha,e/h}(E)$, should satisfy the standard normalization condition
$\psi_{\alpha,e/h}^\dagger(E)\cdot \psi_{\alpha,e/h}^{}(E)=1$.
Since incident states are only defined for $|E_\alpha|>\Delta$, \new{they follow
from Eq.~\eqref{tildepsi} as}
$\psi_{\alpha,e/h}(E)=\sqrt{\frac{2}{1+\rho^2(E_\alpha)}} \,\tilde\psi_{\alpha,e/h}(E)$.

\emph{MAR scattering states.---}We next construct scattering states for the finite-voltage case taking into account MAR processes.  Typical MAR trajectories in energy space (``MAR ladder'') are shown in Fig.~\ref{fig1}.  We consider an incident $\alpha$-mover which is an electron or hole like quasiparticle with energy $E$ in the respective continuum, $|E_\alpha|>\Delta$. 
For each step of the MAR ladder sketched in Fig.~\ref{fig1}, 
the energy of an electron changes by $\pm eV$ for right- or left-movers 
when traversing the normal junction region, and similarly the energy shift for holes is $\mp eV$.
The energy $E_n$ of an outgoing (reflected or transmitted) state may therefore involve the 
emission or absorption of an integer number \new{$n$ of  $eV$ quanta,  $E_n=E+neV$.}  
Noting that there are four possible types of incident states, labeled by $s\in \{1,2,3,4\}$ depending on whether an electron- or a hole-type state is injected from the left or from the right side, we obtain a general \emph{Ansatz}
for  MAR scattering states. For $x=0^\pm$, the corresponding bispinor states have the form 
\begin{eqnarray} \nonumber
\Psi_E(0^-,t) &=&e^{-iEt}  \left( \begin{array}{c} \delta_{s,1}\, \psi_{+,e}(E) \\ \delta_{s,2} \, \psi_{-,h}(E)\end{array} \right) \\
\nonumber &+& \sum_n e^{-iE_n t}\left( \begin{array}{c} a_n\tilde\psi_{+,h}(E_n) \\ b_n \tilde\psi_{-,e}(E_n)
\end{array} \right), \\ \label{MARAnsatz}
\Psi_E(0^+,t) &=&e^{-iEt}  \left( \begin{array}{c} \delta_{s,3} \,\psi_{+,h}(E) \\ \delta_{s,4} \,\psi_{-,e}(E)\end{array} \right)\\
&+&\nonumber \sum_n e^{-iE_n t}
\left( \begin{array}{c} c_n \tilde\psi_{+,e}(E_n) 
\\ d_n \tilde\psi_{-,h}(E_n)\end{array} \right).
\end{eqnarray}
Keeping the incident quasiparticle energy $E$ and scattering channel $s$ implicit,  
the complex-valued scattering amplitudes $(a_n,b_n,c_n,d_n)$ appearing in the outgoing spinor ($\tilde \psi_{\alpha,e/h})$ contributions 
in Eq.~\eqref{MARAnsatz} are determined from the matching conditions in Eq.~\eqref{BC},
\new{cf.~Fig.~\ref{fig1}}.  As a result, the scattering amplitudes
satisfy a set of recurrence relations encoding the MAR ladder, \new{see \cite{PRB}}.  

Given a solution of the recurrence relations, using the Fermi function $n_F(E)=1/(e^{E/T}+1)$ and superconducting density of states factors ($\Theta$ is the Heaviside function),
$\nu_{\alpha=\pm} (E) =  \frac{|E_{\alpha}|}{\sqrt{E_{\alpha}^2-\Delta^2}}
\Theta(|E_{\alpha}|-\Delta),$
the $I$-$V$ characteristics follows as
\begin{equation}\label{MARcur}   
I(V) = \frac{e}{2h} \sum_{\alpha=\pm}\int dE\, n_F(E)\nu_\alpha(E) 
    I_\alpha(r,E)+ (r\to -r),
\end{equation}
with the reflection amplitude $r$  in Eq.~\eqref{BC} and the current matrix elements 
 \begin{eqnarray}\nonumber
I_\alpha (r,E) &=&  \sum_{{\rm odd}\,n} \Bigl [ 
|c^{}_{\alpha,n}|^2 \left(1+|\rho(E+neV-v_Fq)|^2\right) \\
&-&  |d_{\alpha,n}|^2 \left(1+|\rho(E+neV+v_Fq)|^2\right)
 \Bigr]. \label{iam}
 \end{eqnarray}
By taking advantage of symmetry relations connecting the solutions incident from the left side ($s=1,2$) to 
those incident from the right side ($s=3,4$), the latter solutions are contained in Eq.~\eqref{MARcur} through 
the term with $r\to -r$.  The index $\alpha$ in Eq.~\eqref{iam} then corresponds to $s=1$ (for $\alpha=+$) and  $s=2$ (for $\alpha=-$).  
The current expression in Eq.~\eqref{MARcur}  affords a transparent physical interpretation.  Summing over all scattering channels $s$ and integrating
over all incident energies $E$, the
weight of the corresponding incident state in the current is determined by the product of the Fermi function, the density of states, 
and a current matrix element.  The latter follows by summing over all orders $n$ of the MAR ladder, where current contributions
only arise  for odd $n$.  At given order $n$,  electrons $(\propto |c_{\alpha,n}|^2)$ and holes $(\propto |d_{\alpha,n}|^2)$ enter with 
opposite sign, where the corresponding Doppler-shifted energy $E_n\mp v_Fq$ appears in the Andreev reflection amplitude $\rho(E)$.

\new{Below we consider the zero-temperature limit.
For arbitrary system parameters represented by the dimensionless quantities $q\xi$, ${\cal T}$, and $eV/\Delta$,
the  efficiency $\eta(V)$ in Eq.~\eqref{efficiency} follows from Eq.~\eqref{MARcur} 
by numerically solving the recurrence relations which can be truncated 
at order $n_{\rm max}\sim 2\Delta/e|V|$. 
It is thus numerically difficult to reach very low voltages for high transparency, where the MAR ladder in Fig.~\ref{fig1} includes a large number of round trips in the junction region.  Our code accurately reproduces known results for $I_{q=0}(V)$ \cite{Averin1995} and 
for the solvable case ${\cal T}=1$ with $q\ne 0$. }  

\emph{Ballistic limit.---}For ${\cal T}=1$, the matching conditions \eqref{BC} 
as well as the BdG Hamiltonian conserve chirality, $\sigma_z=\alpha=\pm$, and the 
recurrence relations admit a closed solution  \cite{PRB}.
We find that $I(V)$ for $q\ne 0$ is related to the known $q=0$ curve $I_{q=0}(V)$ \cite{Averin1995}  by
a simple shift,
\begin{equation}\label{currdopplershift}
    I(V) = I_{q=0}(V)+ \frac{4e\Delta}{h}  q\xi.  
\end{equation}
This shift has a clear physical interpretation: it is the current carried by Cooper pairs with finite momentum $2q$ and charge $2e$.  The simple decomposition \eqref{currdopplershift} only 
applies in the ballistic limit where chirality is conserved.   
The rectification efficiency  then follows from Eq.~\eqref{efficiency} as
\begin{equation}\label{rectball}
    \eta(V,q\xi,{\cal T}=1) = \frac{4e\Delta}{h} \frac{q\xi}{I_{q=0}(V)},
\end{equation}
\new{which depends on the voltage only through the ratio $eV/\Delta$, i.e., 
the Doppler shifted gaps $\Delta_\pm$ do not appear.
For $e|V|\gg \Delta$, the Ohmic result of a normal-conducting contact, $I_0(V)\approx (2e^2/h)V$, implies $\eta(V) \simeq  2q\xi \frac{\Delta}{eV}$.
On the other hand, for $e|V|\ll \Delta$,} using $I_{q=0}(V\to 0)\approx (4e \Delta/h)\,{\rm sgn}(V)$ \cite{Averin1995}, Eq.~\eqref{rectball} gives $\eta(V)\simeq q\xi.$ 
For $q\xi\to 1$, one approaches the ideal rectification limit since the MAR-induced current $I_0$ now
precisely cancels the finite-momentum Cooper pair current for $V<0$, i.e, $I(V<0)=0$ in Eq.~\eqref{currdopplershift}, while both currents add for $V>0$ to give $I(V>0)=8e\Delta/h$. 
 As a result, we have $\eta(V)=1$. \new{Clearly, there is no current suppression for $q\xi\to 1$ even though one of the spectral gaps vanishes, $\Delta_-\to 0$.}
We conclude that MAR processes can generate highly efficient superconducting diode 
behavior in the deep subgap regime $e|V|\ll \Delta$.

\begin{figure}
\includegraphics[width=0.49\textwidth]{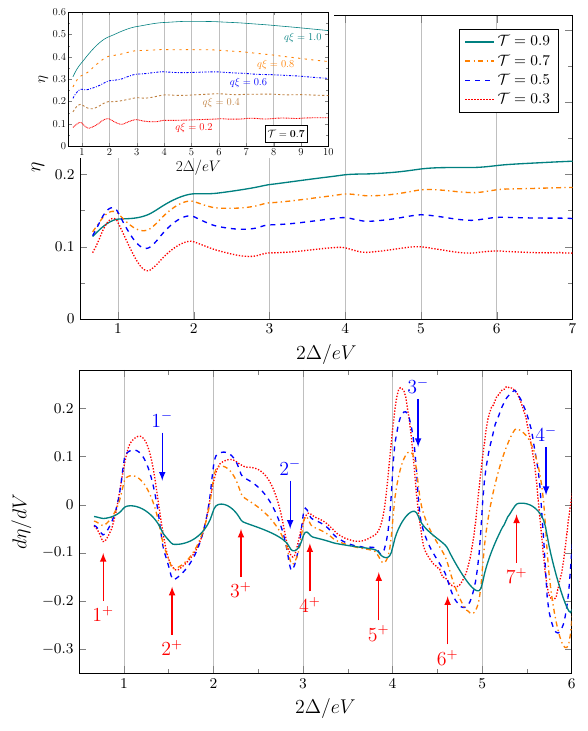}
\caption{Rectification efficiency $\eta(V)$ obtained by numerical evaluation of Eqs.~\eqref{efficiency} and \eqref{MARcur}. 
Top panel: The main part shows $\eta(V)$ vs $2\Delta/eV$ 
for $q\xi=0.3$ and several ${\cal T}$.  Dotted vertical lines indicate standard 
MAR features at $2\Delta/eV=n$ (integer $n$) \cite{Klapwijk1982,Bratus1995,Averin1995,Cuevas1996}.
The inset is for ${\cal T}=0.7$ and different $q\xi$. 
Bottom panel:  $d\eta/dV$ vs $2\Delta/eV$ for $q\xi=0.3$ and the same ${\cal T}$ as in the 
main part of the top panel.
Arrows labeled by $n^\pm$ indicate the points where $2\Delta_\pm/eV=n$. 
}
\label{fig2}
\end{figure}

\begin{figure}
\includegraphics[width=0.47\textwidth]{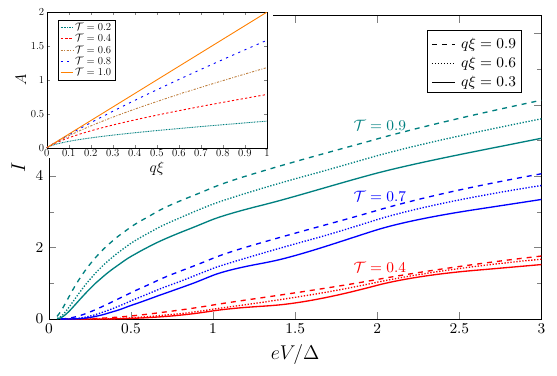}
\caption{\new{$I$-$V$ curves and high-voltage rectification. 
Main panel:  $I$-$V$ curves for different values of $(q\xi,{\cal T})$, with currents in units of $2e\Delta/h$.  Inset: Rectification coefficient $A$ vs $q\xi$ for several ${\cal T}$ in the large-voltage limit, see Eq.~\eqref{Adef}, obtained by numerical solution
of Eq.~\eqref{MARcur} for $eV/\Delta=50$.} 
 }
\label{fig3}
\end{figure}

\emph{Subharmonic structure.---}In Fig.~\ref{fig2}, we show numerical results for 
$\eta(V)$ for different values of $(q\xi,{\cal T})$.  
We observe an overall increase of $\eta(V)$ with increasing Cooper pair 
momentum $2q$ and/or junction transparency ${\cal T}$.
The efficiency is particularly large in the subgap regime $eV\alt 2\Delta$, where we also observe a subharmonic structure with peaks or dips. 
\new{The enhancement of $\eta(V)$ for $e|V|\ll \Delta$  compared to $\eta_0$ 
(for otherwise identical parameters) thus persists for ${\cal T}<1$, 
with optimal rectification at ${\cal T}=1$.
The subharmonic structure is} more clearly visible 
in the derivative $d\eta(V)/dV$ (bottom panel in Fig.~\ref{fig2}).
Apart from the standard $q=0$ MAR features at $2\Delta/eV=n$ (integer $n$),
which are also observed for $q\ne 0$ and follow from MAR trajectories as 
drawn in the upper panel in Fig.~\ref{fig1}, we also find 
resonances or antiresonances corresponding to the Doppler-shifted pairing gaps $\Delta_\pm$ (indicated by arrows in Fig.~\ref{fig2}). 
The corresponding transitions are naturally explained from the MAR ladder picture shown in the lower panel of Fig.~\ref{fig1}, where the presence of normal reflection $r\ne 0$ enables
MAR trajectories between states near the same type of spectral gap ($\pm\Delta_+$ or $\pm\Delta_-$) where $\nu_\alpha(E)$ has sharp peaks. 
\new{In Fig.~\ref{fig3}, we illustrate the $I$-$V$ curves for several
values of $(q\xi,{\cal T})$, see also \cite{PRB}. Note that the current remains large for $q\xi\to 1$.  For $eV\gg \Delta$, we find}
\begin{equation}\label{Adef}
\eta(eV\gg\Delta,q\xi,{\cal T}) \simeq A(q\xi,{\cal T}) \frac{\Delta}{eV}.
\end{equation}
\new{The dimensionless coefficient $A(q\xi,{\cal T})$ is shown
in the inset of} Fig.~\ref{fig3}, with $A=2q\xi$ for ${\cal T}=1$ from the analytical solution. 
For $q\xi \alt 1$, our numerical results suggest $A(q\xi,{\cal T})\approx 2q\xi{\cal T}$.

\emph{Conclusions.---}We have studied voltage-biased Josephson diodes with finite Cooper pair momentum. In the subgap regime, we find that MAR processes allow for large rectification efficiencies \new{accompanied by large currents.  We note that the $I$-$V$ curve for $e|V|\ll \Delta$ can be computed analytically  
from a time average over quasi-stationary Andreev levels \cite{Averin1995,PRB} by using the Josephson relation $\dot\varphi=2eV/\hbar$.  In this case, relaxation mechanisms mixing Andreev states are inefficient and one can obtain perfect rectification ($\eta=1$) at 
the optimal working point $q\xi=1$. On the other hand, under current-biased conditions \cite{Davydova2022}, an effective relaxation mechanism is tacitly assumed, and this results in  $\eta_0\alt 0.4$ and a different optimal working point ($q\xi\approx 0.9$). 
One may expect a similar rectification efficiency enhancement in voltage-biased Josephson diodes
where other mechanisms are responsible for the SDE.} We hope that future experimental and theoretical work will shed light on this intriguing question.

\begin{acknowledgments}
We thank Liang Fu for discussions. 
We acknowledge funding by the Deutsche Forschungsgemeinschaft (DFG, German Research Foundation) under Grant No.~277101999 - TRR 183 (project C01), Grant No.~EG 96/13-1, 
and under Germany's Excellence Strategy - Cluster of Excellence Matter and Light for Quantum Computing (ML4Q) EXC 2004/1 - 390534769.
This work received support from the French government under the France 2030 investment plan, as part of the Initiative d'Excellence d'Aix-Marseille Universit\'e - A*MIDEX, through the institutes IPhU (AMX-19-IET-008) and AMUtech (AMX-19-IET-01X).
\end{acknowledgments}

\bibliography{sup}
\end{document}